\documentclass[12pt]{article}
\usepackage{amsfonts}
\usepackage{amsmath}
\usepackage{amssymb}
\usepackage{eucal}
\numberwithin{equation}{section}

\begin{document}
\title{Exact Solutions of Teukolsky Master Equation with Continuous Spectrum}
\vskip 1.5truecm
\author{Roumen~S.~Borissov \thanks{
E-mail:\,\,\,borissov@phys.uni-sofia.bg, Phone: +359-2-816-1819,
Fax:+359-2-962-4951}, Plamen~P.~Fiziev
\thanks{ E-mail:\,\,\,fiziev@phys.uni-sofia.bg, Phone: +359-2-962-4951,
Fax:+359-2-962-4951}\\ Physics
Department, Sofia University ``St. Kliment Ohridski", \\ 5
James Bourchier Blvd., 1164 Sofia, Bulgaria }
\date{}
\maketitle
\begin{abstract}
Weak gravitational, electromagnetic, neutrino and scalar fields,
considered as perturbations on Kerr background satisfy Teukolsky
Master Equation. The two non-trivial equations obtained after
separating the variables are the polar angle equation and the
radial equation. We solve them by transforming each one into the
form of a confluent Heun equation. The transformation depends on a
set of parameters, which can be chosen in a such a way, so the
resulting angular and radial equations separately have simple
polynomial solutions for neutrino, electromagnetic, and
gravitational perturbations, provided some additional conditions
are satisfied. Remarkably there exists a class of solutions for
which these additional conditions are the same for both the
angular and the radial equations for spins $|s|=1/2$ and $|s|=1$.
As a result the additional conditions fix the dependence of the
separation constant on the angular frequency but the frequency
itself remains unconstrained and belongs to a continuous spectrum.
\end{abstract}

\section{Introduction}
Fields of various types -- scalar, neutrino, electromagnetic, and
gravitational -- have been extensively studied as perturbations to
known solutions of Einstein's equations for configurations with
spherical and cylindrical symmetry. The fields considered are weak
in the sense that we can neglect the influence of their
stress-energy tensor on the background metric. Regge and Wheeler
\cite{ReggeWheeler} and Zerilli \cite{Zerilli,Johnston} were the
first to study the linear response of Schwarzschild solution of
Einstein's equations to perturbations. In order to study Kerr
metric perturbations Teukolsky \cite{Teukolsky}-\cite{Teukolsky3}
analyzed the components of Weyl tensor, using Newman-Penrose
formalism \cite{NP}. (For a detailed extended presentation see
\cite{Chandra}.). As a result one obtains the Teukolsky Master
Equation, which describes the dynamics of various fields of
different spins as perturbations to Kerr metric. In recent years
there is an increased interest on the subject
\cite{Andersson}-\cite{Hod}, mostly aimed at studying the
quasi-normal modes. Another problem analyzed via Teukolsky's
equations is related to the decaying of the various fields present
during a gravitational collapse at very late times at large
distances -- the so called late-time tails. All these
investigations however, are performed via indirect, approximate
methods \cite{Andersson,Kokkotas},\cite{Hod}-\cite{Leaver1}.

On the other hand, already for quite some time it has been
recognized in the literature \cite{Marc1,Marc2,GE,STU} that
Regge-Wheeler and Teukolsky's equations can be transformed into
the form of a confluent Heun equation \cite{Heun}-\cite{Fiziev6}. The
reason there has not been much attention paid to the Heun-type
solutions is that they are not completely analyzed and, in
general, difficult to work with. Some basic classes of exact
solutions to Rege-Wheeler equation in terms of special solutions
to the confluent Heun equation -- the so-called confluent Heun
functions (see the Appendix), were described recently and were
used for finding solutions to a number of physical problems
\cite{Fiziev1}-\cite{Fiziev3}.

Following the articles \cite{Fiziev1}-\cite{Fiziev3} we continue
with the application of the confluent Heun functions to
Teukolsky's equations. The first results, presented in
\cite{PF1}-\cite{PFDS4}, were very encouraging  and drew special
attention to the solutions in terms of the confluent Heun
polynomials \cite{Heun}-\cite{Fiziev6} (see the Appendix). It should
be emphasized that long time ago in \cite{Marc1,Marc2} it was
recognized by Baldin, Pons, and Marcilhacy that the conditions for
polynomial solutions to Heun equations lead to polynomial
solutions to Teukolsky's equations in a generalized sense,
i.e. polynomials multiplied by non-polynomial factors which are elementary functions.
Having in mind the general
description of all 256 classes of factorized solutions to
Teukolsky Master Equation \cite{Fiziev4,Fiziev5} we intend to focus on the
mathematical properties of some of them and study various physical
applications.

The general description of all polynomial solutions of Teukolsky
Master Equation was given for the first time in  \cite{Fiziev4, Fiziev5}.
These fall into two different classes. For the first class, the
first polynomial condition (\ref{alphaint}), called the
$\delta_N$-condition in \cite{Fiziev6}, \cite{Fiziev4,Fiziev5}, is automatically
satisfied. For waves of spin $|s|$ this condition fixes only the
degree $(N+1)=2|s|$ of the second polynomial condition
$\Delta_{N+1}=0$.
For the second class of polynomial solutions the
$\delta_N$-condition is fulfilled only for certain complex
frequencies $\omega_N$ which belong to definite equidistant
discrete spectra. For the two classes the second polynomial
condition $\Delta_{N+1}=0$ defines an algebraic equation of degree
$2|s|$ for the second separation constant: $E_m=E_m(\omega)$, and
$E_m=E_m(\omega_N)$, correspondingly.

Here we are considering only polynomial solutions of the first
class. Thus an independent derivation of the specific relations,
valid only for the first class of polynomial solutions becomes
possible. It is based on a direct check of the two necessary
conditions (\ref{alphaint}) and (\ref{conditpoly}), which together
are sufficient to ensure the polynomial character of the solutions
(See the Appendix.).

Below we present an independent derivation of the first class
polynomial solution both for Teukolsky's angular and radial
equations using the notations of reference \cite{Ron}. This
notation has some advantages since it simplifies significantly the
form of the $\delta_N$-condition. The correspondence between the
notation of \cite{Ron} and the notation used in \cite{DDLMRR1,DDLMRR2},
\cite{Fiziev6},\cite{Fiziev1}-\cite{Fiziev5} and in the computer
application Maple is described in section 9.4 of the Appendix.

In the present paper we consider a specific type of evolution of
week fields with spin $|s|=1/2, 1\,\,\text{and}\,\, 2$ on Kerr
background. The solutions studied here are double polynomial
solutions that describe one-way waves of corresponding spins, the
so-called total transmission modes. These are factorized solutions
to Teukolsky Master Equation, in which the solutions both of the
angular and the radial equations (of the same spin weight $s$)
belong to the corresponding first classes of polynomial solutions,
introduced in \cite{Fiziev4,Fiziev5}. Here we show that  these solutions
yield a complex one-parameter continuous spectrum of the frequency
$\omega$ and derive the explicit form of the separation constant
$E$ in the various cases. Finally we discus some overall solutions
of Teukolsky Master Equation, constructed making use only of these
continuous spectrum solutions.

To the best of our knowledge this is the first time when
for Teukolsky Master Equation exact solutions with continuous
spectrum  are presented for a specific boundary problem.
An interesting observation is that continuous
spectrum emerges only for neutrino and electromagnetic waves,
because of the simultaneous fulfillment of the polynomial
condition both for the angular and the radial Teukolsky equations.
We have to stress that such a simultaneous fulfillment is not in
place for gravitational waves. The physical consequences of this
mathematical result may be deep and very important. Its roots can
be traced back to some results, originally obtained in
\cite{Teukolsky3} and developed further in \cite{Chandra}. We
present here the mathematical basis, needed for further
developments in this direction.

In the next section we start by reminding the procedure for
separation of the variables in Teukolsky Master Equation via
factorization of the solutions and the corresponding  basic
results. In section 3 we present the general scheme for
transforming Teukolsky's radial equation (TRE) into the one of the
many known "canonical" forms of the confluent Heun equation
\cite{Heun}-\cite{Fiziev6}\footnote{In the literature there still
does not exist a commonly accepted standard form both of the five
classes of Heun equations and of the corresponding Heun functions.
Different possible forms are in use, since in different
applications the authors prefer the form, which is most suitable
for their specific needs.}, namely into the so-called
non-symmetrical canonical form. We show that for specific values
of the indices of the regular singular points \cite{Ince} the
first condition for polynomial solution to TRE in the form of a
confluent Heun equation is automatically attained. We impose the
second condition for having a polynomial solution and obtain the
value of the separation parameter $E$ as function of the frequency
$\omega$. In section 4 we continue by presenting the
transformation to non-symmetrical canonical Heun form of
Teukolky's angular equation (TAE) and again show that for
corresponding specific choice of the indices of its regular
singular points we obtain again a polynomial solution. In order to
achieve this result we derive the explicit form of the second
polynomial condition and arrive at the result that in some cases
it is the same as for the radial equation. Thus we find a
simultaneous fulfillment of the polynomial conditions for the
angular and for the radial Teukolsky's equations for perturbations
with spin $|s|= 1/2$ and $1$. In section 5 we discuss why such  a
simultaneous fulfillment of the polynomial conditions is not
possible for gravitational waves ($|s|=2$). In section 6 we
present the basic properties of the overall solutions to the
Teukolsky Master Equation constructed only from the factorized
solutions with continuous spectrum. We show that these solutions
describe one-way collimated waves, which may be regular along the
rotational axes, despite the singular character of the polynomial
solutions of the angular Teukolsky equation. In mathematical sense
these solutions form a natural orthogonal basis of singular
functions for integral representation of physically meaningful
solutions. In the conclusion we give a brief summary and ideas for
future studies on the matter.

In an Appendix some basic properties of the confluent Heun
equation and its solutions and different forms are presented for
the reader's convenience.

\section{Spin weight $s$ fields on Kerr background}
In this section we present some basic results of Teukolsky's
approach \cite{Teukolsky}-\cite{Teukolsky3} to the perturbations
of spin $|s|$ of Kerr vacuum solution for the metric of a rotating
black hole. In Boyer-Lindquist coordinates the metric is given by
\cite{Chandra}, \cite{MTW}:

\begin{eqnarray}\label{KNMETRIC}
ds^2 &=& {\left( 1- \frac{2Mr}{\Sigma} \right)} dt^2 + \frac{{
4aMr\sin^2\theta }}{\Sigma} dt d\phi - { {\Sigma} \over {\Delta} }
dr^2\nonumber \\ &&-\,\Sigma d\theta^2 -
{\left[r^2+a^2+\frac{2Ma^2r\sin^2\theta}{\Sigma}\right]}
\sin^2\theta d\phi^2\,.
\end{eqnarray}
Here $M$ is the Keplerian mass of the rotating black hole and $a$
is its angular momentum per unit mass. Also, ${\Delta}$ and
${\Sigma}$ are defined in the usual way:
\begin{eqnarray}\label{DEL}
\Delta \equiv r^2 - 2Mr + a^2, \qquad \Sigma \equiv r^2 +
a^2\cos^2\theta\,.
\end{eqnarray}
The dynamics of a massless field $\Psi = \Psi(t,r,\theta,\phi)$
with spin weight {\it s} is described by Teukolsky Master
Equation:

\begin{eqnarray}\nonumber
&& \left[\frac{\left(\ r^2+a^2 \right)^2}{\Delta} - a^2
\sin^2\theta \right]{\partial^2\Psi \over \partial t^2} + {\frac
{4Mar}{\Delta}}{\partial^2\Psi \over \partial t
\partial\phi} + {\left[ {a^2 \over \Delta} - {1 \over
\sin^2\theta }\right]}{\partial^2\Psi \over \partial\phi^2} -
\\ \nonumber
&& - \Delta^{-s}\frac{\partial}{\partial
r}\left(\Delta^{s+1}\frac{\partial\Psi}{\partial r}\right) - {1
\over \sin \theta} {\partial \over \partial \theta} \left(\sin
\theta {\partial\Psi \over \partial \theta}\right)- 2s{\left[
\frac{a\left( r-M\right)}{\Delta} +
\frac{i\cos\theta}{\sin^2\theta}
\right]\frac{\partial\Psi}{\partial\phi}} - \\ \nonumber \\ && -
2s{\left[ \frac{M\left( r^2-a^2\right)}{\Delta} - r -ia\cos\theta
\right]\frac{\partial\Psi}{\partial t}} + \left( s^2\cot^2\theta -
s \right)\Psi =0\,.\label{TME}
\end{eqnarray}
In the above equation we have the following expressions for
$\Psi$:
\begin{itemize}
\item For $s=1/2$, $\Psi=\chi_{0}$ and for $s=-1/2$,
$\Psi=\rho^{-1}\chi_{1}$, where $\chi_{0}$ and $\chi_{1}$
represent the two components of the neutrino spinor in
Newman-Penrose formalism. \item For $s=1$, $\Psi=\varphi_{0}$ and
for $s=-1$, $\Psi=\rho^{-2}\varphi_{2}$, where $\varphi_{0}$ and
$\varphi_{2}$ are Maxwell tensor tetrad components in
Newman-Penrose formalism. \item In the gravitational case for
$s=2$, $\Psi=\psi_{0}$ and for $s=-2$, $\Psi=\rho^{-4}\psi_{4}$,
where $\psi_{0}$ and $\psi_{4}$ are Weyl tensor Newman-Penrose
tetrad components.
\end{itemize}
In all cases $\rho=-1/(r-ia\cos\theta )$ and the choice of the
Kinnersley tetrad is assumed. In order to separate the variables
following \cite{Teukolsky}-\cite{Teukolsky3} we set

\begin{equation}\label{Fourier}
\Psi(t,r,\theta,\phi)=  {1 \over 2\pi} \int_{-\infty +
i\epsilon}^{\infty +i\epsilon} \sum_{m=-\infty}^{\infty}
\Psi_{m}(\omega; t,r,\theta,\phi) d\omega\,,
\end{equation}
where $\Psi_{m}(\omega; t,r,\theta,\phi)= e^{im\phi} e^{-i\omega
t} S_{m}(\omega;\theta)R_{m}(\omega;r)$, and $\epsilon$ is a
parameter defining the contour of integration. The azimuthal
number has values $m = 0,\pm 1,\pm 2, \dots$ for integer spin, or
$m = \pm 1/2,\pm 3/2, \dots$ for half-integer spin
\cite{Gold,Dolan1,Dolan2,Fiziev5}.

Thus we are looking for an integral representation with a {\em
factorized} kernel $\Psi_{m}(\omega; t,r,\theta,\phi)$. For the
unknown factors $S_{m}(\omega;\theta)$ and $R_{m}(\omega;r)$ we
obtain respectively Teukolsky's angular equation (TAE):
\begin{eqnarray}\label{angular}
&& {1 \over \sin \theta}{\partial \over \partial\theta} \left(
\sin \theta \frac{\partial }{\partial\theta}
\right)S_{m}(\omega;\theta) + \\ \nonumber && +
\left(a^2\omega^2\cos^2\theta - {2a\omega s \cos\theta}-
\frac{m^2+s^2+2ms\cos\theta}{\sin^2\theta} + E_{m}
\right)S_{m}(\omega;\theta) = 0\,,
\end{eqnarray}
and Teukolsky's radial equation (TRE)
\begin{equation}\label{radial}
\Delta^{-s} {d \over dr} \left( \Delta^{s+1} \frac{d }{d r}
\right)R_{m}(\omega;r)+ \left(\frac{K^2-2is\left( r-M
\right)K}{\Delta} + 4is\omega r - \lambda_{m}
\right)R_{m}(\omega;r) = 0\,,
\end{equation}
where $K \equiv \left( r^2 +a^2 \right)\omega - am$ and
$\lambda_{m} \equiv E_{m} + a^2\omega^2-2am\omega -s(s+1)$ is the
separation ``constant"\footnote {In the literature often another
form of the separation constant is used, namely
$A_{m}=E_{m}-s(s+1)$ so $\lambda_{m}$ can also be written as
$\lambda_{m} \equiv A_{m} + a^2\omega^2-2am\omega $.}.  These are
the two equations, some special solutions of which we will study in detail in our paper.

\section{Solutions to TRE}
\subsection{Transforming TRE into the
non-symmetrical canonical form of Heun equation}

Both equations \eqref{angular} and \eqref{radial} are second order
ordinary differential equations with two regular singular points
at finite value of the independent variable and one irregular
singularity at infinity. The most general equation with such
properties is the confluent Heun equation \cite{Ron}. Thus both
TRE and TAE are specific cases of confluent Heun equations. At
this point we continue by transforming the radial equation
\eqref{radial} into one of the forms of the confluent Heun
equation, the non-symmetrical canonical form \eqref{H}. In order
to do so we apply the so-called {\it s}-homotopic transformation
to \eqref{radial} by setting\footnote{This anzatz was used for the
first time for analytical and numerical studies of the problem at
hand by Leaver \cite{Leaver1}, \cite{Leaver, Leaver2}.}
\begin{equation}\label{anzatzR}
R_{m}(\omega;r) = (r-r_{+})^{\xi}(r-r_{-})^{\eta}e^{\zeta
r}H(r)\,,
\end{equation}
where $r_{+}$ and $r_{-}$ are the event and Cauchy horizons of a
Kerr black hole defined by $r_{\pm} = M \pm \sqrt{M^2 - a^2}$ and
$\xi$, $\eta$ (the indices of the regular singularities) and
$\zeta$ are parameters to be determined. By substituting
\eqref{anzatzR} into \eqref{radial} and after some straightforward
algebra we arrive at the following equation for $H(r)$:

\begin{eqnarray}\nonumber
&&{d^2 H(r) \over dr^2} + \left( \frac{2\xi+s+1}{r-r_{+}} +
\frac{2\eta+s+1}{r-r_{-}} +2\zeta\right) {d H(r) \over dr} +
\\ \nonumber \\ \nonumber &&\frac{1}{(r-r_{+})(r-r_{-})}\{
\left[4\omega^2 M + 2\zeta(\xi+\eta + s+1)+2is\omega \right]r +
\\ \nonumber \\ \nonumber &&+(\xi + \eta)^2 + (\xi +
\eta)+2s(\xi+\eta)-2\zeta (\xi r_{-} + \eta r_{+}) - 2s\zeta M
-2\zeta M- \\ \nonumber \\ \nonumber &&-2a\omega m +4\omega^2 M^2
-2is\omega M - \lambda_{m} \}H(r) = 0\,.
\end{eqnarray}
In order to obtain this form of the equation we had to fix the
parameters $\xi$, $\eta$ and $\zeta$. For them we end up with
quadratic equations and thus with pairs of possible expressions,
namely:

\begin{eqnarray}\label{valuesxi} \xi_{1} =
ia_{+} (\omega - m\Omega_{+}) , \qquad \xi_{2} = -s - ia_{+}
(\omega - m\Omega_{+})\,,
\end{eqnarray}

\begin{eqnarray}\label{valueseta} \eta_{1} = -s +
ia_{-} (\omega - m\Omega_{-}), \qquad \eta_{2} = - ia_{-} (\omega
- m\Omega_{-})\,,
\end{eqnarray}
and

\begin{eqnarray}\label{valueszeta} \zeta_{1} =  i\omega ,
\qquad \zeta_{2} = - i\omega\,,
\end{eqnarray}
where we have set

\begin{eqnarray}\label{defaOmega}
a_{\pm}= {2Mr_{\pm} \over r_{+} - r_{-}}, \qquad \Omega_{\pm}= {a
\over 2Mr_{\pm} } \,.
\end{eqnarray}
Any one of the eight possible triplets of expressions from
\eqref{valuesxi}, \eqref{valueseta}, and \eqref{valueszeta} leads
to a Heun equation in the desired form with different parameters.
In order to determine these parameters it is necessary to perform
one more step and to introduce dimensionless variables instead of
$r$. In order to keep the symmetries in the problem manifest it is
best to set different variables for positive and for negative spin
weights, namely for  $s=1/2, 1, 2$ and for $s=-1/2, -1, -2$ we
will have respectively:
\begin{eqnarray}\nonumber
{}_{+}z=\frac{r_{+}-r}{r_{+}-r_{-}}, \qquad
{}_{-}z=\frac{r-r_{-}}{r_{+}-r_{-}}.
\end{eqnarray}
Note that from now on in the paper we will denote the sign of the
spin weight (and when there might be a confusion the spin weight
itself) by a subscript to the left of the variable. Also always
the signs "+", "-", or $\pm$ used as subscripts to the right from
the variables will be related to the two singularities in the
radial equation, namely to the event and the Cauchy horizons.
After performing these adjustments we end up with a Heun equation
in the non-symmetrical standard form \eqref{H}:
\begin{eqnarray} \label{eqHrlast}
{{d^2H}\over{d({}_{\pm}z)^2}}+\left(4({}_{\pm}p)+{{{}_{\pm}\gamma}\over{{}_{\pm}z}}+
{{{}_{\pm}\delta}\over{{}_{\pm}z-1}}\right){{dH}\over{d({}_{\pm}z)}}+{4({}_{\pm}\alpha)
({}_{\pm}p)({}_{\pm}z) -({}_{\pm}\sigma)
\over{({}_{\pm}z)(({}_{\pm}z)-1)}}H=0
\end{eqnarray}
with the parameters given by the following expressions:
${}_{\pm}p= \mp (r_{+}-r_{-}) \zeta/2$, ${}_{+}\gamma =
{}_{-}\delta = 2\xi +s +1$, ${}_{-}\gamma = {}_{+}\delta = 2\eta +
s + 1$,

\begin{eqnarray}\label{valuesHalpha}
{}_{\pm}\alpha = 2M\omega^2 \zeta^{-1} +\xi + \eta + s + 1 +
is\omega \zeta^{-1}\,,
\end{eqnarray}
and
\begin{eqnarray}\nonumber
&& {}_{\pm}\sigma = - 2\zeta r_{\pm}  \left( {2M\omega^2 \over
\zeta} +\xi + \eta + s + 1 + {is\omega \over \zeta} \right) - [
(\xi + \eta)^2 + (\xi + \eta)+ 2s(\xi+\eta) -   \\ \nonumber \\
\nonumber && \qquad - 2\zeta (\xi r_{-} + \eta r_{+}) - 2s\zeta M
-2\zeta M -2a\omega m +4\omega^2 M^2 -2is\omega M - \lambda_{m}]
\,,
\end{eqnarray}
for any choice of a triplet $\xi, \eta, \zeta$. In general there
exist pairs of solutions to Heun equation expressed as appropriate
power series about each one of the singularities. The solutions of
the equation \eqref{eqHrlast} can be expressed with the use of the
so-called Frobenius solutions
$Hc^{(r)}({}_{\pm}p,{}_{\pm}\alpha,{}_{\pm}\gamma,{}_{\pm}\delta,
{}_{\pm}\sigma;{}_{\pm}z)$. There exists a multitude of other
solutions to Heun equation which can be obtained  from these
Frobenius solution by interchanging the finite singular points or
performing appropriate {\it s}-homotopic transformations (See the
Apendix).

\subsection{Polynomial solutions}
Thus we have completed the transformation of TRE to the
non-symmetrical form of the confluent Heun equation and we can
formally identify the solutions which are given by specific Heun
functions. There exists though a special case, in which the
confluent Heun equation admits polynomial solutions (See the
Appendix). This special case depends on the values of
${}_{\pm}\alpha$ in \eqref{valuesHalpha} and another, more
involved condition on the parameters in the Heun equation. If the
parameters ${}_{\pm}\alpha$ are equal to negative integer numbers
or zero and if we can solve the second condition then we will have
polynomial solutions to Heun equation. At this point we can make
the following observation regarding the possible values of
${}_{\pm}\alpha$ in \eqref{valuesHalpha}: In the cases of positive
spin weights if we pick the values $\xi_{2}$, $\eta_{1}$, and
$\zeta_{2}$ from \eqref{valuesxi}, \eqref{valueseta}, and
\eqref{valueszeta}, then we obtain ${}_{+}\alpha=1-2s$, which
equals $0$ for neutrino, $-1$ in the electromagnetic case and $-3$
in the gravitational case. Similarly, for $s=-1/2$, $s=-1$, or
$s=-2$ we have to choose $\xi_{1}$, $\eta_{2}$, and $\zeta_{1}$
and we will get ${}_{-}\alpha=1+2s$, which again gives
$\alpha=0,-1$ or $-3$ for the neutrino, for the electromagnetic,
and for gravitational case respectively. Thus we have identified
particular combinations of the values of the parameters $\xi$,
$\eta$, and $\zeta$ for which we may expect to find polynomial
solutions of the radial equation for $|s|=1/2, 1, 2$ since the
necessary condition $\alpha=-N$, $N$-integer, is satisfied. Also,
it can be seen easily that for $s=0$ we obtain ${}_{\pm}\alpha=1$
so there is no polynomial solution for scalar fields. The second
condition will depend on the value of $N$. It is an algebraic
equation of order $N+1$ and leads to polynomial solutions to Heun
equations of order $N$. Thus we expect that in the neutrino case
the polynomial solutions, if they exist, are simply constants, in
the electromagnetic case - linear functions, and in the
gravitational case the solutions are cubic polynomials. We will
consider here in more detail only the electromagnetic case. The
results for $s=\pm 1/2$ can be easily obtained using the same
procedure. The polynomial solutions to TRE with spin $|s|=2$ were
described long time ago in quite a different setting by
Chandrasekhar \cite{Chandra84}. For a more recent treatment see
\cite{Brink}.

\subsection{Polynomial solution to TRE for spin $|s|= 1$}

In order to find a polynomial solution to TRE we have to impose in
addition to the fact that ${}_{\pm}\alpha=-N$ the condition
\eqref{conditpoly}. In this case we are looking for an expansion
about the singular point at infinity. The second polynomial
condition for electromagnetic perturbations has the form
$g^{(r)}_{0}g^{(r)}_{1}=h^{(r)}_{1}f^{(r)}_{0}$ with the
coefficients from the three-term relation \eqref{3termrelation2}
given by $g^{(r)}_{0}=-\sigma -4p+\gamma+\delta,
g^{(r)}_{1}=-\sigma, f^{(r)}_{0}= -4p, h^{(r)}_{1}= -\gamma$.
Provided we make the above mentioned choices rendering
${}_{\pm}\alpha=-1$ we obtain the following values for the
remaining parameters in the Heun equation: ${}_{\pm}p = (i/2)
\omega (r_{+}-r_{-}), {}_{\pm}\gamma = \pm 2i a_{-} (\omega -
m\Omega_{-}), {}_{\pm}\delta = \mp 2ia_{+} (\omega - m\Omega_{+}),
{}_{\pm}\sigma = {}_{\pm}E_{m} +a^2\omega^2 -2a\omega m \mp
2i\omega r_{\pm} $. This second polynomiality condition
essentially fixes ${}_{\pm}\sigma$. It leads to a quadratic
equation for ${}_{\pm}\sigma$ with solutions ${}_{\pm} \sigma_{1}
= \mp 2i\omega r_{\pm} + 2\sqrt{a\omega (a\omega -m)}$ and
${}_{\pm} \sigma_{2} = \mp 2i\omega r_{\pm} - 2\sqrt{a\omega
(a\omega -m)}$ (recall that the $\pm$ sign to the left of $\sigma$
refers to the sign of the spin weight, while the subscripts $1$
and $2$ number the solutions to the quadratic equation).

The first important result from solving the second condition for
having polynomial solutions is that we find the dependence of the
separation ``constant" $E_{m}$ on the frequency $\omega$, which is
the same for both $s=+1$ and for $s=-1$ \cite{Fiziev4,Fiziev5} and is given  by
\begin{eqnarray}
&& \nonumber {}_{\pm1} E_{m}(a\omega)_{1} = -a^2\omega^2 +
2a\omega m + 2\sqrt{a\omega (a\omega -m)}
\\ && \nonumber {}_{\pm1} E_{m}(a\omega)_{2} = -a^2\omega^2 + 2a\omega m - 2\sqrt{a\omega
(a\omega -m)} \,.
\end{eqnarray}

By returning to the $r$ variable we arrive at the following
expressions for the polynomial solutions of the Heun equation for
$s=1$:

\begin{eqnarray}\label{Hemplusr12}
({}_{+1}H_{m}(\omega;r))_{1,2}={1 \over r_{+}-r_{-}} \left(- r \pm
{i \over \omega} \sqrt{a\omega (a\omega - m)}\right)  \,.
\end{eqnarray}
Similarly, for $s=-1$ we have:
\begin{eqnarray}\label{Hemminusr12}
({}_{-1}H_{m}(\omega;r))_{1,2}={1 \over r_{+}-r_{-}} \left(r \pm
{i \over \omega} \sqrt{a\omega (a\omega - m)}\right)\,.
\end{eqnarray}
Putting together \eqref{anzatzR} with the specific values of
$\xi$, $\eta$, and $\zeta$ and the polynomial solutions of Heun
equations, the solutions to the TRE for $s=1$ and for $s=-1$ can
be written (modulo normalizing constants) respectively as:

\begin{eqnarray}\label{finalRplus}
&& ({}_{+1}R_{m}(\omega;r))_{1,2} =
 {e^{-i\omega r_{*}} \over \Delta}
{\left({r-r_{+} \over r-r_{-}}\right)^{{ima \over r_{+} - r_{-}}}}
 ({}_{+1}H_{m}(\omega;r))_{1,2}
\,,
\end{eqnarray}
and
\begin{eqnarray}\label{finalRminus}
&& ({}_{-1}R_{m}(\omega;r))_{1,2} =
 e^{i\omega r_{*}}
{\left({r-r_{+} \over r-r_{-}}\right)^{-{ ima \over r_{+} -
r_{-}}}}
 ({}_{-1}H_{m}(\omega;r))_{1,2}
\,,
\end{eqnarray}
where
\begin{equation}\nonumber
r_{*}=r + a_{+}\ln|r-r_{+}| - a_{-}\ln|r-r_{-}|
\end{equation}
is the ``tortoise" coordinate and $a_{\pm}$ are defined in
\eqref{defaOmega}. Note that these exact solutions for $|s|=1$ are
presented here for the first time in explicit form. A similar form
of polynomial solutions to TRE in the case $|s|=2$ can be found in
\cite{Brink}.

Using the orthogonality relations of Heun
polynomials \cite{Ron} it can be shown that in terms of the intermediate
variables ${}_{s}z = {}_{\pm}z$ we have:

\begin{equation}\label{ortho}
\int_{-\infty}^{0}\Delta^{s}({}_{s}R_{m}(\omega;{}_{s}z))_{j}
({}_{s}R_{m}(\omega;{}_{s}z))_{l} d({}_{s}z) =0 , \qquad j \neq l
\qquad j,l=1,2 \,,
\end{equation}
where $\Delta$ is the standard factor from the Kerr metric,
defined in \eqref{DEL}. The behavior of these solutions at
infinity and at the event horizon can be readily determined form
\eqref{finalRplus} and \eqref{finalRminus}. First, both solutions
in \eqref{finalRplus} with $s=1$ have the behavior:

\begin{eqnarray}\label{Hemplusinfty}
({}_{+1}R_{m}(\omega;r))_{1,2} \sim  \left\{
\begin{array}{lcr}
r^{-1}e^{-i\omega r_{*}} \; & r \rightarrow \infty & (r_{*}
\rightarrow \infty) \\
\\ \Delta^{-s} e^{-i\varpi r_{*}} \; & r \rightarrow r_{+}
& (r_{*} \rightarrow -\infty)
\end{array}\right.,
\end{eqnarray}
while both solutions in \eqref{finalRminus} with $s=-1$ behave
like

\begin{eqnarray}\label{Hemminusinfty}
({}_{-1}R_{m}(\omega;r))_{1,2} \sim  \left\{
\begin{array}{lcr}
 r^{-(2s+1)} e^{i\omega r_{*}}  \; & r \rightarrow \infty & (r_{*} \rightarrow
\infty) \\
\\e^{i\varpi r_{*}}  \; & r \rightarrow r_{+}
& (r_{*} \rightarrow -\infty)
\end{array}\right.,
\end{eqnarray}
where in the expressions for the behavior at $r_{+}$ we have set
$\varpi = \omega - m\Omega_{+}$. Thus we found exact solutions to
TRE, the nature of which depends on the relative sign between
$\omega$ and $\varpi$, in agreement with the general analysis in
\cite{Teukolsky}-\cite{Teukolsky2}. When $\omega$ and $\varpi$
have the same sign then the solutions we found describe one-way
waves traveling from $r_{+}$ to infinity or in the opposite
direction. The solutions with $\omega$ and $\varpi$ with opposite
signs describe either waves traveling towards $r_{+}$ and towards
infinity or leaving from $r_{+}$ and coming from infinity.

\subsection{Polynomial solution for spin $|s|= {1 / 2}$}

This case follows along the same lines like the electromagnetic
perturbations but is simpler since $\alpha=0$. This leads to
$c^{(r)}_{1}=0$ in \eqref{3termrelation2} which corresponds to a
constant solution to the Heun equation. This implies that
$\sigma=0$ and thus determines the dependence of the separation
constant $E$ from $\omega$. The result is \cite{Fiziev4,Fiziev5}

\begin{equation}\nonumber
{}_{\pm {1 \over2}}E_{m}(a\omega)= - a^2\omega^2 + 2a\omega m - {1
\over 4}.
\end{equation}
The solutions to TRE for $s=\pm {1 \over 2}$ are presented here for the first time:

\begin{equation} \label{Rneuplus}
 {}_{1 \over 2}R_{m}(\omega;r) =  {e^{-i\omega r_{*}} \over
 \sqrt{\Delta}}
{\left({r-r_{+} \over r-r_{-}}\right)^{{ima \over r_{+} - r_{-}}}}
 {}_{1 \over 2} H_{m}(\omega)\,
\end{equation}

\begin{equation} \label{Rneuminus}
{}_{-{1 \over 2}}R_{m}(\omega;r) =  e^{i\omega r_{*}}
{\left({r-r_{+} \over r-r_{-}}\right)^{-{ima \over r_{+} -
r_{-}}}} {}_{-{1 \over 2}}H_{m}(\omega)\,,
\end{equation}
where ${}_{1 \over 2}H_{m}(\omega)$ and ${}_{-{1 \over
2}}H_{m}(\omega)$ are constants in $r$. Note that these solutions
satisfy the same orthogonality relations as those in \eqref{ortho}
and have the same behavior at infinity and at $r_{+}$ as those
given in \eqref{Hemplusinfty} and \eqref{Hemminusinfty}.

\section{Solutions to TAE}

\subsection{Transformation of TAE into the form of a Heun equation}

In order to transform TAE \eqref{angular} into the form of a Heun
equation we follow the same procedure as with the radial one.
Following \cite{Fackerell}-\cite{Berti} we set

\begin{equation}\label{substang}
 S_{m}(\omega;u) =
(1-u)^{\mu_{1}}(1+u)^{\mu_{2}}e^{\nu u}T_{m}(\omega;u)\,,
\end{equation}
where $u =\cos\theta$ and $\mu_{1}$ and $\mu_{2}$ are the indices
of the regular singular points at $\theta=0$ and $\theta=\pi$. We
plug \eqref{substang} into \eqref{angular} and obtain an equation
for $T_{m}(\omega;u)$. Imposing the condition that the equation
for $T_{m}(\omega;u)$ has the form \eqref{H} leads to a system of
quadratic algebraic equations for the parameters $\mu_{1}$,
$\mu_{2}$, and $\nu$. Their solutions are given by
\begin{eqnarray}\label{uvwvalues}
\mu_{1}=\pm \frac{m+s}{2} , \qquad \mu_{2}=\pm \frac{m-s}{2}, \qquad \nu=\pm
a\omega \,.
\end{eqnarray}
For these values of $\mu_{1}$, $\mu_{2}$, and $\nu$ the equation
for $T_{m}(\omega;u)$ is given by

\begin{eqnarray}\label{eqHa}
&&{d^2 T_m(\omega;u) \over du^2}+ \left( \frac{2\mu_{1}+1}{u-1} +
\frac{2\mu_{2}+1}{u+1} +2\nu \right) {d T_m(\omega;u) \over du} +
\\ \nonumber && \frac{1}{(u-1)(u+1)}\{ 2\nu \left(
\mu_{1}+\mu_{2}+1+{a\omega s \over \nu }\right)u - \\ \nonumber \\
\nonumber && - \left[ E_{m} + a^2\omega^2 -(\mu_{1}+\mu_{2})^2 -
(\mu_{1}+\mu_{2}) -2\nu ( \mu_{1}-\mu_{2})\right]\}T_m(\omega;u) =
0 \,.
\end{eqnarray}
In order to complete the transformation we introduce new
independent variables (the reason for the specific form of this
transformation will become clear latter) $u \mapsto {}_{+}x$ for
$s=+1/2,1,2$ and $u \mapsto {}_{-}x$ for $s=-1/2,-1,-2$, where
${}_{\pm}x=(1\pm u)/2$. It is important to notice that these two
independent variables are related by the transformation $\theta
\mapsto \pi-\theta$. Thus we arrive at the following specific
non-symmetric canonical form of a confluent Heun equation:
\begin{equation}\label{eqHalast}
{d^2 T_m(\omega;{}_{\pm}x) \over d{({}_{\pm}x)}^2}+ \left(
\frac{{}_{\pm}\gamma}{{{}_{\pm}x}} +
\frac{{}_{\pm}\delta}{{{}_{\pm}x}-1} +4({}_{\pm}p) \right) {d
T_m(\omega;{{}_{\pm}x}) \over d({}_{\pm}x)} +
 \frac{4({}_{\pm}p)({}_{\pm}\alpha) ({}_{\pm}x) -
({}_{\pm}\sigma)}{{{}_{\pm}x}({{}_{\pm}x}-1)}T_m(\omega;{{}_{\pm}x})
= 0 \,,
\end{equation}
with the identification $ {}_{\pm}p= \pm \nu$, ${}_{+}\gamma =
{}_{-}\delta = 2\mu_{2}+1 $, ${}_{-}\gamma = {}_{+}\delta =
2\mu_{1}+1$, and for both signs we have

\begin{eqnarray}\nonumber
 && {}_{\pm} \alpha =
\mu_{1}+\mu_{2}+1+{a\omega s \over \nu } \\ \nonumber \\
\label{paramH} && {}_{\pm} \sigma = - 2\nu ({}_{\pm} \alpha) +
E_{m} + a^2\omega^2 -(\mu_{1}+\mu_{2})^2 - (\mu_{1}+\mu_{2}) -2\nu
(\mu_{1}-\mu_{2}) \,.
\end{eqnarray}
The solutions of equation \eqref{eqHalast} can be expressed with
the use of the Frobenius solutions
$Hc^{(a)}(p,\alpha,\gamma,\delta,\sigma;{}_{\pm}x)$ and all other
solutions obtained from it by interchanging the finite singular
points or performing appropriate {\it s}-homotopic transformations
(See the Apendix). Thus we obtained the needed solutions to the
Heun equation obtained from TAE\footnote{The  Tom\'{e} asymptotic
solution at infinity $Hc^{(r)}(p,\alpha,\gamma,\delta,\sigma;x)$
is not of physical significance in the case of angular equation
(\ref{eqHa}), because here we are interested only in solutions on
the interval $u\in (-1,1)$.}. In the general case these are given
by two sets of infinite series about each one of the regular
singularities. For more detailed description of all local
solutions to the angular Teukolsky equation see \cite{Fiziev4,Fiziev5}.
With these we can build the solutions to TAE.

For each choice of a triple $\mu_{1}$, $\mu_{2}$, and $\nu$ we get
a different solution to TAE. There are two special cases though:
First, we can choose $\mu_{1}$, $\mu_{2}$, and $\nu$ in such a way
so to obtain solutions that are regular at both $\theta=0$ and
$\theta=\pi$. This chice leads to a well studied Sturm-Liouville
eigenvalue problem \cite{Fackerell}-\cite{Berti} for the
spin-weighted spheroidal wave functions $S_{lm}(\omega;\theta)$
\cite{AS} and the separation constant $E_{lm}=E_{lm}(a\omega)$,
which for fixed $s$, $m$, and $a\omega$ are labeled by an
additional integer $l$. The eigenfunctions $S_{lm}(\omega;\theta)$
are complete and orthogonal on $0\leq \theta \leq \pi$ for each
set $s$, $m$, and $a\omega$. In the case $s=0$,
$S_{lm}(\omega;\theta)$ are the spheroidal wave functions. When
$a\omega=0$, the eigenfunctions are the spin-weighted spherical
harmonics ${}_{s}Y^{m}_{l}={}_{s}S^{m}_{l}(\theta)e^{im\phi}$. In
the general case, as it is shown for example in \cite{Seidel} the
function $T_m(\omega;z)$ can be expanded as a series of Jacobi
polynomials in the case of an integer spin, and to their
spin-weighted generalizations for half-integer spins
\cite{Gold,Dolan1,Dolan2}, and one can obtain the separation
constant $\lambda_{lm}$ as a power series in $a\omega$.

The second special case corresponds to polynomial solutions to TAE
\cite{Fiziev4,Fiziev5}. Let us choose $\mu_{1}= -(m+s)/2,
\mu_{2}=(m-s)/2, \nu=  - a\omega$. This choice leads in
\eqref{paramH} to $\alpha = 1-2s$ , which for $s=1/2$, $s=1$ and
$s=2$ is zero or a negative integer number. This means that the
condition \eqref{alphaint}, which is necessary for having a
polynomial solution (a constant in the case $\alpha=0$) of the
confluent Heun equation is met for these values of the spin weight
$s$. The case of zero spin is again excluded by this condition. If
instead we have $s=-1/2$, $s=-1$ or $s=-2$, we must choose
$\mu_{1}= (m+s)/2, \mu_{2}= -(m-s)/2, \nu= a\omega$, and this will
give us $\alpha = 2s+1$, which in this case again is zero or a
negative integer number. We have to emphasize that this result
comes at a price - depending on the specific values of $m$ and $s$
either one or the other of the pre-factors in \eqref{substang}
diverge at the corresponding singular point. This means that for
the cases of neutrino, electromagnetic and gravitational
perturbations we eventually may write down the solutions to TAE
into the form of diverging at some of the singularities pre-factor
multiplied by a polynomial expression, simultaneously regular at
both regular singular points. Again, the neutrino case is
relatively simple and can be easily deduced from the
electromagnetic one, which we will present in detail.

\subsection{Polynomial solutions for perturbations with spin $|s|= 1$}

Since \eqref{alphaint} is satisfied we can obtain a polynomial
solution of \eqref{eqHalast} by imposing as an additional (already
sufficiency) condition \eqref{conditpoly}. In our case
\eqref{conditpoly} translates into an $\omega$ dependence of the
separation constant $E_{m}$, which enters in the parameter
$\sigma$ from \eqref{paramH}.

In this case $\alpha=-1$, so we will have $c^{(a)}_{2}=0$  in
\eqref{conditpoly}. Because of the specific choice we made when
introducing ${}_{\pm}x$ in this case we have for both positive and
negative spin weights the same values of the parameters in the
Heun equation: $p= -a\omega, \alpha=-1, \gamma = m, \delta = -m,
\sigma = E_{m} + a^2\omega^2 +2a\omega -2a\omega m$. This means
that we obtain the same Heun equation for both $s=1$ and $s=-1$
with Frobenius solutions
$Hc^{(a)}(p,\alpha,\gamma,\delta,\sigma;{}_{\pm}x)$. The only
difference between the two cases is in the definition of the
independent variable ${}_{\pm}x=(1\pm \cos\theta)/2$. Thus in this
section we will look for a solution only for the case $s=1$ and
will obtain the solution for $s=-1$ by performing in the final
results the transformation $\theta \mapsto \pi - \theta$. In terms
of the coefficients from the three term relation
\eqref{3termrelation1} the sufficiency condition for $m\neq 0$ has
the form $g^{(a)}_{0}g^{(a)}_{1}=h^{(a)}_{1}f^{(a)}_{0}$, where
$g^{(a)}_{0}=-\sigma, g^{(a)}_{1}= 4a\omega -\sigma, f^{(a)}_{0}=
- m, h^{(a)}_{1}= 4a\omega $. Since this condition leads to a
quadratic equation for $ \sigma$ we obtain pairs of solutions. For
both $s=+1$ and $s=-1$ we get the same expressions $\sigma_{1} =
2a\omega + 2\sqrt{a\omega (a\omega -m)}$ and $\sigma_{2} =
2a\omega - 2\sqrt{a\omega (a\omega -m)}$. Returning back to the
original variables of the angular equation we can write down the
following expressions \cite{PFDS3,PFDS4},  for the solutions for
$s=+1$ (for $m \neq 0$):

\begin{equation} \label{emSplusfinal}
({}_{+1}S_{m}(\omega;\theta))_{1,2} ={e^{-a\omega \cos\theta}
\over \sin\theta} \left(\cot{\theta \over 2}\right)^{m}
({}_{+1}T_{m}(\omega;\theta))_{1,2}\,,
\end{equation}
and we have restored the pre-subscripts denoting the two different
spin weights. The polynomial parts of the solutions for $m \neq 0$
are:
\begin{equation} \nonumber
({}_{+1}T_{m}(\omega;\theta))_{1} = 1 - \frac{2a\omega +
2\sqrt{a\omega(a\omega - m)}}{m}\cos^{2}{\theta \over 2}\,,
\end{equation}

\begin{equation} \nonumber
({}_{+1}T_{m}(\omega;\theta))_{2} = 1 - \frac{2a\omega -
2\sqrt{a\omega(a\omega - m)}}{m}\cos^{2}{\theta \over 2} \,.
\end{equation}
In the case $m=0$ it is easy to solve the equation directly and
arrive at the following overall solutions:
\begin{eqnarray} \nonumber
({}_{+1}S_{0}(\omega;\theta))_{1} =
e^{-a\omega\cos\theta}\tan{\theta \over 2}, \qquad
({}_{+1}S_{0}(\omega;\theta))_{2} =
e^{-a\omega\cos\theta}\cot{\theta \over 2} \,.
\end{eqnarray}
The behavior of the solutions to TAE we found at the two
singularities $\theta=0$ and $\theta=\pi$ can be easily obtained
from the expressions above. Each one is divergent either at the
one or at the other singularity. The exact expressions for $m \neq
0$ are:

\begin{eqnarray}\label{Semplus}
({}_{+1}S_{m}(\omega;\theta))_{1,2} \sim  \left\{
\begin{array}{lcr}
 \theta^{-(m+1)}  &  \mbox{at}\;\theta \rightarrow 0 &
\mbox{for}\; m\geq1 \\ \\ (\pi - \theta)^{-(m+1)} &
\mbox{at}\;\theta \rightarrow \pi & \mbox{for}\; m\leq1
\end{array}\right..
\end{eqnarray}
For $m=0$ we have $({}_{+1}S_{0}(\omega;\theta))_{1} \sim
\theta^{-1}$ at $\theta \rightarrow 0$ and
$({}_{+1}S_{0}(\omega;\theta))_{2} \sim (\pi - \theta)^{-1}$ at
$\theta \rightarrow \pi$. If we perform the transformation $\theta
\mapsto \pi - \theta$ we obtain the solutions for $s=-1$:

\begin{equation} \label{emSminusfinal}
({}_{-1}S_{m}(\omega;\theta))_{1,2} = {e^{a\omega \cos\theta}
\over \sin\theta} \left(\cot{(\pi - \theta) \over 2}\right)^{m}
({}_{-1}T_{m}(\omega;\theta))_{1,2}\,,
\end{equation}
where for $m \neq 0$:

\begin{equation} \nonumber
({}_{-1}T_{m}(\omega;\theta))_{1} = 1 - \frac{2a\omega +
2\sqrt{a\omega(a\omega - m)}}{m}\sin^{2}{\theta \over 2}\,,
\end{equation}

\begin{equation} \nonumber
({}_{-1}T_{m}(\omega;\theta))_{2} = 1 - \frac{2a\omega -
2\sqrt{a\omega(a\omega - m)}}{m}\sin^{2}{\theta \over 2}\,,
\end{equation}
and for $m=0$:

\begin{eqnarray} \nonumber
({}_{-1}S_{0}(\omega;\theta))_{1} =
e^{a\omega\cos\theta}\cot{\theta \over 2}, \qquad
({}_{-1}S_{0}(\omega;\theta))_{2} =
e^{a\omega\cos\theta}\tan{\theta \over 2} \,.
\end{eqnarray}
It can be shown that because of the orthogonality properties of
Heun polynomials \cite{Ron}, the following relations hold for the functions
$({}_{\pm1}S_{m}(\omega;x))_{j}$ separately for $s=1$ and for
$s=-1$:

\begin{eqnarray} \label{ortho2}
\int_{0}^{1}
({}_{\pm1}S_{m}(\omega;{}_{\pm1}x))_{j}({}_{\pm1}S_{m}(\omega;{}_{\pm1}x))_{l}
d({}_{\pm1}x) = 0 \, \qquad j \neq l \, \qquad j,l=1,2.
\end{eqnarray}

The second condition \eqref{conditpoly} for a polynomial solution
of the radial equation leads to the expressions for the separation
``constant" $E_{m}$  for both $s=+1$ and for $s=-1$. For any $m$
we have \cite{Fiziev4,Fiziev5}
\begin{eqnarray}
&& \nonumber {}_{\pm1} E_{m}(a\omega)_{1} = -a^2\omega^2 +
2a\omega m + 2\sqrt{a\omega (a\omega -m)}
\\ && \nonumber {}_{\pm1} E_{m}(a\omega)_{2} = -a^2\omega^2 + 2a\omega m - 2\sqrt{a\omega
(a\omega -m)} \,.
\end{eqnarray}
Surprisingly, these expressions are the same both for the  TRE and
TAE (see Section 3.2).

Hence, the second condition \eqref{conditpoly} does not produce an
additional relation between $E_{m}$ and $\omega$. The main
consequence from this result is that we can express the separation
constant $E_{m}$ as a function of $\omega$ but the (complex)
frequency itself remains unconstrained. As a result we obtain a
continuous spectrum in $\omega$ for the solutions of Teukolsky's
angular and radial equations \eqref{angular} and \eqref{radial}.

Thus, the surprising phenomenon of simultaneous fulfilment of the
polynomial conditions for angular and radial equations with spin
$|s|=1$ is related to the existence of the continuous spectrum of
Teukolsky Master Equation \eqref{TME} in the specific boundary
problem under consideration.

\subsection{Polynomial solutions for perturbations with spin
$|s|= {1/2}$}

Again we have $\sigma=0$ and this determines the dependence of the
separation constant $E$ from $\omega$. The result again is \cite{Fiziev4,Fiziev5}

\begin{equation}\nonumber
{}_{\pm {1 \over2}}E_{m}(a\omega)= - a^2\omega^2 + 2a\omega m - {1
\over 4}.
\end{equation}
Like in the electromagnetic case both conditions for polynomiality
are the same. Thus in this case again we obtain a continuous
spectrum. The solutions to TAE for $s=\pm {1 \over 2}$, presented here for the first time, are:

\begin{equation} \label{Sneuplus}
{}_{ {1 \over2}} S_{m}(\omega;\theta) = \left( \cot{\theta \over
2} \right)^{m} {e^{-a\omega \cos\theta} \over \sqrt{\sin\theta}}
\,,
\end{equation}

\begin{equation} \label{Sneuminus}
{}_{-{1 \over2}}S_{m}(\omega;\theta) = \left( \tan{\theta \over 2}
\right)^{m} {e^{a\omega \cos\theta} \over \sqrt{\sin\theta}}\,.
\end{equation}
These solutions satisfy the same orthogonality relations as in
\eqref{ortho2}. The behaviour at $\theta = 0$ and at $\theta =
\pi$ can be easily deduced from \eqref{Sneuplus} and
\eqref{Sneuminus}.

\section{Nonexistence of gravitational one-way waves of continuous spectrum}

The derivation of the polynomial solutions for perturbation with spin $|s|=2$
can be found in \cite{Chandra84,Brink}.
Our analysis in the gravitational case follows the same lines as in the electromagnetic one
and reproduces the results of these articles.
As it was already discussed, with appropriate choice of the powers in \eqref{anzatzR} and
\eqref{substang} we achieve $\alpha=-3$ for both TRE and TAE. When
we impose the second polynomiality condition we obtain quartic
equations for the values of the separation ``constant". The
important difference is that unlike in the electromagnetic and the
neutrino cases, the Heun polynomial conditions are different for
the angular and for the radial equations. The proof based on Taylor series expansions
of the roots of quartic equations can be found in \cite{Fiziev4,Fiziev5}.
This result is consistent with the observation by Teukolsky and Press \cite{Teukolsky2},
developed further by Chandrasekar \cite{Chandra},
that the difference between the Starobinsky's constant for the
angular and for the radial equations in the gravitational case is
equal to $(12M\omega)^2$. Thus in the gravitational case we have
two independent conditions, relating $E_{m}$ and $\omega$ which
leads to a discrete spectrum of $\omega$ labeled by $m$
and some additional indexes for the different solutions of the quartic
equations.

\section{Overall solutions to Teukolsky's Master Equation for spin $|s|= 1$}

\subsection{Polynomial in both $r$ and $\cos\theta$, diverging
at $\theta=0$ and $\theta=\pi$ solutions}

At this point we are ready to return to the original physical
fields. For $s=1$ and $s=-1$ we have respectively
$\Psi=\varphi_{0}$ and $\Psi=\rho^{-2}\varphi_{2}$. With the
solutions we found we can write:

\begin{eqnarray}\nonumber
&& ((\varphi_{0})_{m}(\omega;t,r,\theta,\phi))_{1,2} \sim
{e^{-a\omega \cos\theta} e^{im\phi} \over \sin\theta}
\left(\cot{\theta \over 2}\right)^{m}  \times \\ \nonumber
\\ && \nonumber \qquad \times  {e^{-i\omega (r_{*}+t)} \over \Delta}
\exp{\left({i m a \over r_{+} - r_{-}}\ln\left\arrowvert{r-r_{+}
\over r-r_{-}}\right\arrowvert\right)}
({}_{+}T_{m}(\omega;\theta))_{1,2} ({}_{+}H_{m}(\omega;r))_{1,2}
\,,
\end{eqnarray}
and
\begin{eqnarray}\nonumber
&& ((\varphi_{2})_{m}(\omega;t,r,\theta,\phi))_{1,2} \sim
{e^{a\omega \cos\theta}e^{im\phi} \over \sin\theta}
\left(\tan{\theta \over 2}\right)^{m}  \times \\ \nonumber
\\ && \nonumber \qquad \times {e^{i\omega (r_{*}-t)} \over
(r-ia\cos\theta)^2}
\exp{\left(-{ima \over r_{+} - r_{-}}\ln\left\arrowvert{r-r_{+}
\over r-r_{-}}\right\arrowvert\right)}
({}_{-}T_{m}(\omega;\theta))_{1,2} ({}_{-}H_{m}(\omega;r))_{1,2}
\,.
\end{eqnarray}
These expressions can be written in a more compact form if we
introduce the Kerr coordinates with the relations:

\begin{eqnarray}\nonumber
&& \tilde{V}=t+r_{*}, \qquad \qquad \qquad \qquad \qquad \qquad
\tilde{U}=t-r_{*} \\ \nonumber && {}_{+}\tilde{\phi}=\phi+{a \over
r_{+} - r_{-}}\ln\left\arrowvert{r-r_{+} \over
r-r_{-}}\right\arrowvert, \qquad {}_{-}\tilde{\phi}=\phi-{a \over
r_{+} - r_{-}}\ln\left\arrowvert{r-r_{+} \over
r-r_{-}}\right\arrowvert \,.
\end{eqnarray}
Thus the final expressions for the solutions to Teukolsky Master
Equation will be

\begin{equation}\label{finaloverall}
((\varphi_{0})_{m}(\omega;t,r,\theta,\phi))_{1,2} \sim
{e^{-i\omega \tilde{V}} e^{-a\omega\cos\theta} \over (r^2-2Mr+a^2)
\sin\theta} ({}_{+}W)^{m} ({}_{+}T_{m}(\omega;\theta))_{1,2}
({}_{+}H_{m}(\omega;r))_{1,2}\,
\end{equation}
and
\begin{equation} \nonumber
((\varphi_{2})_{m}(\omega;t,r,\theta,\phi))_{1,2} \sim
{e^{-i\omega \tilde{U}}e^{a\omega\cos\theta} \over
(r-ia\cos\theta)^2 \sin\theta} ({}_{-}W)^{m}
({}_{-}T_{m}(\omega;\theta))_{1,2} ({}_{-}H_{m}(\omega;r))_{1,2}
\,,
\end{equation}
where we have introduced the following expressions
\begin{eqnarray}\label{Wpm}
{}_{+}W =e^{i({}_{+}\tilde{\phi})} \cot\left({{}_{+}\theta / 2}\right), \qquad
{}_{-}W = e^{i({}_{-}\tilde{\phi})} \cot\left({{{}_{-}\theta}/ 2}\right)
\,
\end{eqnarray}
to denote the stereographic projections of a unit sphere
parameterized by angles ${}_{+}\tilde{\phi},\,
{}_{+}\theta=\theta$ for $s=1$ and ${}_{-}\tilde{\phi},\,
{}_{-}\theta=\pi - \theta$ for $s=-1$ on the complex planes
$\mathbb{}{C}_{{}_\pm W}$. The first formula in \eqref{Wpm}
describes stereographic projection from the North pole and the
second one -- from the South one. Note that after the transition
from real variables $({}_{\pm}\theta, {}_{\pm}\tilde{\phi})$ to
the complex one ${}_{+}W$ one must introduce an additional phase
factor $\exp{(-is{}_{\pm}\tilde{\phi})}$ in the spin-weighted
spheroidal harmonics, due to the back rotation of the basis (See
the paper by Goldberg et al. in \cite{Gold}.). In the case of spin
$1/2$ the introduction of such a factor $\exp{(\mp {i \over 2
}{}_{\pm}\tilde{\phi})}$ is equivalent to a transition in what
follows from half-integer to integer values of the azimuthal
number $m$ and a replacement $m \rightarrow \pm 1/2$.

The two expressions in
\eqref{finaloverall} together with their behavior at $r
\rightarrow \infty $ and at the regular singularities of TRE and
TAE provide us with a basis from which we can build solutions with
specific boundary conditions. The basis describes wave collimated
along the poles $\theta=0$ and $\theta=\pi$. Depending on the
values of the parameters $\omega$ and $m$, when $\omega$ and
$\varpi = \omega - m\Omega_{+}$ have the same sign, the basis
describes waves having the same direction of propagation at both
$r_{+}$ and at infinity. Otherwise, when $\omega$ and $\varpi =
\omega - m\Omega_{+}$ have opposite signs, the waves have opposite
directions of propagation at $r_{+}$ and at infinity. One possible
application of this basis is to try to explain the Central Engine
of the Gamma Ray Bursts (GRB), discussed in \cite{PFDS3} and
\cite{PFDS4}.

\subsection{Polynomial in $r$ regular
at $\theta=0$ and $\theta=\pi$ solutions}

We could combine the polynomial solutions to TRE with the
spin-weighted spheroidal harmonics representing the regular at
$\theta=0$ and $\theta=\pi$ solution. The result will be a basis
of waves which have the same properties in radial direction as
those discussed above but will not be collimated along the axes of
rotation. Possible application of the basis in this form is to
study the influence of rotating gravitational field for formation
and evolution of the Supernovae outbursts \cite{Fiziev4,Fiziev5}.

\section{Overall solutions to Teukolsky Master Equation for $|s|=1/2$, diverging
at $\theta=0$ and $\theta=\pi$}

Combining the results from \eqref{Sneuplus}, \eqref{Sneuminus},
\eqref{Rneuplus}, and \eqref{Rneuminus} we can write down the
expressions for the neutrino components $\chi_{0}$ and $\chi_{1}$
as:

\begin{equation}\label{neufinal}
(\chi_{0})_{m}(\omega;t,r,\theta,\phi) \sim { e^{-i\omega
\tilde{V}} e^{-a\omega\cos\theta} \over \sqrt{r^2-2Mr+a^2}
\sqrt{\sin \theta}}
 ({}_{+}W)^{m} \,,
\end{equation}

\begin{equation}\nonumber
(\chi_{1})_{m}(\omega;t,r,\theta,\phi) \sim { e^{-i\omega
\tilde{U}} e^{a\omega\cos\theta} \over (r-ia\cos\theta) \sqrt{\sin
\theta}}
 ({}_{-}W)^{m} \,.
\end{equation}
Using these solutions, which to the best of our knowledge are
published for the first time, we will show how we can build
regular with respect to the $\theta$ solutions. The general
form of the one-way solutions, based on \eqref{neufinal} is

\begin{equation}\label{neusum}
\chi_{0}(\omega;t,r,\theta,\phi) = { e^{-i\omega \tilde{V}}
e^{-a\omega\cos\theta} \over \sqrt{r^2-2Mr+a^2} \sqrt{\sin
\theta}} \sum_{m=-\infty}^{\infty} A_{m}(\omega)
 ({}_{+}W)^{m} \,.
\end{equation}
The physical model is determined by the amplitudes
$A_{m}(\omega)$. Physically sound solutions correspond to
amplitudes $A_{m}(\omega)$, which lead to a finite result after
performing the summation. Here we consider only a formal example
proving the existence of proper choice of the amplitudes giving
finite results. Let us write down the part of $\chi_{0}$
containing the potentially singular factor as
\begin{equation}
(\sin\theta)^{-1/2}\sum_{m=-\infty}^{\infty} A_{m}(\omega)
\left({}_{+}W\right)^{m}=\sqrt{{1 \over 2}\left(\left|{}_{+}W\right|
+  \left|{}_{+}W\right|^{-1}\right)}
\sum_{m=-\infty}^{\infty} A_{m}(\omega) \left({}_{+}W\right)^{m}\,,
\end{equation}
where $\left|{}_{+}W\right|=\cot{\theta \over 2}\geq 0$. At this
point we proceed by choosing in an appropriate way functions
$f(\omega, {}_{+}W)=\sum_{m=-\infty}^{\infty} A_{m}(\omega)
\left({}_{+}W\right)^{m}$ thus defining the amplitudes. For
example we can choose $f(\omega, {}_{+}W)=\big({}_{+}W +
({}_{+}W)^{-1} + \text{const}\big)^{-1}$ or more generally
$f(\omega,{}_{+}W)=\big(P(\omega,{}_{+}W) +
Q(\omega,{}_{+}W^{-1})\big)^{-1}$, where $P$ and $Q$ are arbitrary
polynomials of degree not less than one. Then the amplitudes
$A_{m}(\omega)$ are the coefficients in the Laurent series of the
functions $f(\omega,{}_{+}W)$ with respect to ${}_{+}W$. With this
choice it is clear that there are no singularities in the limits
$\left|{}_{+}W\right| \to 0$ and $\left|{}_{+}W\right| \to
\infty$. It can be shown that choosing the polynomials $P$ and $Q$
properly we preserve the collimation of the neutrino waves.

We can go one step further and formally perform the integration in
\eqref{Fourier}. Thus we arrive at the following general solutions
to Teukolsky's Master Equation for neutrino waves with spin
weights $s=1/2$ and $s=-1/2$ respectively:

\begin{equation}\nonumber
\chi_{0}(t,r,\theta,\phi) = { F_{0}(\tilde{V}-ia\cos\theta,
{}_{+}W) \over \sqrt{\Delta} \sqrt{\sin \theta}} \,,
\end{equation}

\begin{equation}\nonumber
\chi_{1}(t,r,\theta,\phi) = { F_{1}(\tilde{U}+ia\cos\theta,
{}_{-}W) \over \sqrt{\sin \theta}} \,,
\end{equation}
where $F_{0}$ and $ F_{1}$ are arbitrary functions of their
respective variables. The fact that arbitrary $F_{0}$ and $ F_{1}$
satisfy Teukolsky's Master Equation can be verified directly. The
exact forms of  $F_{0}$ and $ F_{1}$ are to be determined by the
specific physical situations. The above formal examples
demonstrate mathematical technics, which make possible the
application of singular solutions to Teukolsky's Master Equation
for description of physical reality.

\section{Conclusion}

In the paper we presented an approach for solving Teukolsky Master
Equation based on the use of the confluent Heun equation. After
separating the variables we showed that both TRE and TAE can be
transformed into the non-symmetric canonical form of the confluent
Heun equation. The transformation depends on a set of parameters
which when properly chosen lead to polynomial solutions to both
Heun equations related to TRE and to TAE. The surprising result
(which simply means that there should be a deeper physical
explanation we do not understand yet) is that for neutrino and for
electromagnetic perturbations we find solutions which have
continuous spectrum, but for gravitational perturbations this
does not happen. The richness of the results we found give us the
opportunity to construct different types of solutions in
accordance with the specific boundary problem we want to study.

There are many possible directions we intend to pursue. One
possibility is to see if indeed we can explain some of the basic
features of the GRB's and of Supernovae using the basis we found.
Related problem is to investigate further the stability of Kerr
black holes.

\vskip 1truecm
{\em \bf Acknowledgements} \vskip .3truecm

The authors would like to thank Dimo Arnaudov and Denitsa Staicova
for valuable comments and suggestions.

P.F. is grateful  to Professor Saul Teukolsky for his comments on the problems,
related with present article, and especially for raising the question about the use
of singular solutions of Teukolsky's Master Equation.

This article was partially supported by the Foundation
"Theoretical and Computational Physics and Astrophysics" and by
the Bulgarian National Scientific Fund under contracts DO-1-895
and DO-02-136.

\section{Author Contributions}

In 2008 R.B. joined the comprehensive program for of the use of
Heun's functions, developed by P.F. since 2005. He made
calculations for Teukolsky radial and angular equations for
electromagnetic case (i.e. for spin-weight $1$) in the notations
of reference \cite{Ron}, which simplify the polynomial conditions
and discovered that these conditions coincide for both Teukolsky
radial and angular equations. R.B. confirmed the results for
spin-weight $1/2$, obtained previously by P.F. in Maple-notation
and wrote down the overall solutions in section 7. The text of the
present article was written by R.B.

P.F. developed the program for of the use of Heun's functions for solution of Regge-Wheeler and
Teukolsky equations since 2005. He found all specific basic classes of solutions to these equations,
in particular,
all polynomial solutions for different integer and half-integer spin-weights,
as well as justification of some properties of the confluent Heun functions.
P.F. explained the relation of the coinciding polynomial conditions
for electromagnetic and neutrino perturbations with a novel continuous
spectrum of Teukolsky master equation and discovered the method of deriving regular solutions of this
equation using proper superposition of the singular polynomial solutions.
He formulated the possible astrophysical applications of the obtained mathematical results.
P.F. is responsible for the references, some corrections and editing of the text.

\section{Appendix: Heun Equation and Heun Functions}
In this appendix we remind the reader some basic information about
the confluent Heun equation and its solutions, following the
notations of the reference \cite{Ron}.

In subsection 9.1 we give a brief description of the non-symmetrical canonical form of
confluent Heun equation and its solutions.
In subsection 9.2 a basic information about local Frobenius and Tom\'{e} solutions around regular
and irregular singular points is presented.
In subsection 9.3 we remind the basic information about polynomial solutions in notations of
\cite{Ron}.
In subsection 9.4 we describe the correspondence between these notations and the notations,
used in the basic articles \cite{DDLMRR1}, \cite{DDLMRR2} on the modern general theory
of all kinds of Heun equations and the properties of their solutions.
The last notations are used in \cite{Fiziev6}, \cite{Fiziev1}-\cite{Fiziev5}.
At present these conventions become more popular, because they are used in the
computer package Maple, based on the articles \cite{DDLMRR1}, \cite{DDLMRR2}.
This package is still the only one for analytical and numerical computer
calculations with Heun equations and Heun functions.
In Maple's Help one can find an available and rich collection of relations
and properties of Heun functions of all kinds.

\subsection{Non-symmetrical canonical form of confluent Heun equation and its solutions}

The general Heun equation is a second order ODE of Fuchsian type
with four regular singular points. In the present paper we have to
solve the confluent Heun equation (CHE) for different cases. It is
relatively well studied \cite{Heun}-\cite{Fiziev6}, but there still
exist essential gaps in the theory. CHE can be obtained from the
general Heun equation by coalescing of two of the singular points
by redefining certain parameters and taking the appropriate
limits. In this way two regular singular points coalesce into one
irregular (in general) point. The solutions of the confluent Heun
equation are relatively well-studied special functions, already
included in modern computer package Maple. These functions
represent non-trivial generalization of known hypergeometric
functions, yet have richer properties, because confluent Heun
equation has one more singular point than the hypergeometric. One
of the canonical forms of the confluent Heun equation is the so
called non-symmetric canonical form \cite{Ron}:
\begin{eqnarray}
{{d^2H}\over{dz^2}}+\left(4p+{{\gamma}\over{z}}+
{{\delta}\over{z-1}}\right){{dH}\over{dz}}+{4\alpha pz -\sigma
\over{z(z-1)}}H=0.\label{H}
\end{eqnarray}
The only regular Frobenius' type solution to \eqref{H} about the
regular singular point $z=0$ is denoted by
$Hc^{(a)}(p,\alpha,\gamma,\delta,\sigma;z)$. It is defined for
non-integral values of $(1-\gamma)$ in the domain $|z|<1$ by the
condition
\begin{equation}\label{Hca}
Hc^{(a)}(p,\alpha,\gamma,\delta,\sigma;0)=1.
\end{equation}
In \cite{Ron} it is called the "angular solution" of the confluent Heun
equation.

Another solution, is the Tom\'{e}'s type asymptotical solution
$Hc^{(r)}(p,\alpha,\gamma,\delta,\sigma;z)$.
It is defined for complex $p=|p|e^{i\varphi}$ in the domain $|z|>1$ by
the condition:
\begin{equation}\label{ Hcr}
\lim_{|z|\rightarrow \infty}
{z^{\alpha}Hc^{(r)}(p,\alpha,\gamma,\delta,\sigma; -|z| e^{-i\varphi})}=1.
\end{equation}
In \cite{Ron} it is called the "radial solution"  of the confluent Heun
equation.

Different pairs of local solutions can be constructed using the combinations of
four known independent transformations of the parameters,
which preserve the chosen canonical form of the Heun Equation.
For example by interchanging the regular singular points $z_{1}=0$ and $z_{2}=1$:
\begin{equation}\nonumber
z \mapsto 1-z,
\end{equation}
one obtains the following new solutions:
\begin{eqnarray} \nonumber
&& Hc^{(a)}(-p,\alpha,\delta,\gamma,\sigma+4p\alpha;1-z) \\ &&
Hc^{(r)}(-p,\alpha,\delta,\gamma,\sigma+4p\alpha;1-z)\,.
\end{eqnarray}

All possible sets of local solutions to Regge-Wheeler and Teukolsky equations
were described for the first time in \cite{Fiziev4,Fiziev5}.

\subsection{Power-series solutions of the confluent Heun equation}

\subsubsection{Taylor series expansion about the regular singularity $z=0$}
If we expand the solution
$Hc^{(a)}(p,\alpha,\gamma,\delta,\sigma;z)$ as a power series

\begin{equation}\label{series}
Hc^{(a)}(p,\alpha,\gamma,\delta,\sigma;z)=\sum_{k=0}^{\infty}c^{(a)}_{k}z^k
\end{equation}
then we get a three-term recurrence relation for the coefficients
$c^{(a)}_{k}$:

\begin{eqnarray}\label{3termrelation1}
&& f^{(a)}_{k}c^{(a)}_{k+1}+g^{(a)}_{k}c^{(a)}_{k}+h^{(a)}_{k}c^{(a)}_{k-1}=0
\\ \nonumber && c_{-1}=0, \qquad c_{0}=1\,,
\end{eqnarray}
where

\begin{eqnarray}\nonumber
&& g^{(a)}_{k}=k(k-4p+\gamma+\delta-1)-\sigma \\ \nonumber &&
f^{(a)}_{k}=-(k+1)(k+\gamma) \\ \nonumber &&
h^{(a)}_{k}=4p(k+\alpha-1) \,.
\end{eqnarray}

The radius of convergence of the series \eqref{series} is equal to
unity, which is the distance to the next regular singular point
\cite{Ron}.

\subsubsection{Laurent series expansion about the singular point at infinity}

Another power series can be constructed at infinity. In general
this series is not convergent but only asymptotic. For the
function $Hc^{(r)}(p,\alpha,\gamma,\delta,\sigma;z)$ we will have
the expansion

\begin{equation}\label{seriesinf}
Hc^{(r)}(p,\alpha,\gamma,\delta,\sigma;z)=z^{-\alpha}\sum_{k=0}^{\infty}c^{(r)}_{k}z^{-k}.
\end{equation}
The three-term recurrence relation for the coefficients
$c^{(r)}_{k}$ reads

\begin{eqnarray}\label{3termrelation2} &&
f^{(r)}_{k}c^{(r)}_{k+1}+g^{(r)}_{k}c^{(r)}_{k}+h^{(r)}_{k}c^{(r)}_{k-1}=0
\\ \nonumber && c_{-1}=0, \qquad c_{0}=1\,,
\end{eqnarray}
with the following expressions\footnote{Note that these
expressions are somewhat different from those in \cite{Ron}.} for
$f^{(r)}_{k}$, $g^{(r)}_{k}$, and $h^{(r)}_{k}$:

\begin{eqnarray}\nonumber
&& g^{(r)}_{k}=(\alpha+k)(\alpha+k+4p-\gamma-\delta+1)-\sigma \\
\nonumber &&  f^{(r)}_{k}=-4p(k+1) \\ \nonumber &&
h^{(r)}_{k}=-(k+\alpha-1)(\alpha+k-\gamma) \,.
\end{eqnarray}
It is easy to show that in general the series expansion
\eqref{seriesinf} diverges \cite{Ron}.

\subsection{Polynomial solutions of the confluent Heun equation}

Let us consider the case in which the parameter $\alpha$ has a
fixed negative integral value

\begin{equation}\label{alphaint}
\alpha = - N,\qquad N\in \mathbb{N}.
\end{equation}
In this case the coefficient $h^{(a)}_{N+1}$ vanishes for both
expansions above. If we impose in addition the second condition
that

\begin{equation}\label{conditpoly}
c^{(a)}_{N+1}=0
\end{equation}
then the recurrence relation breaks down and we obtain a
polynomial of $N$-th order instead of the infinite series. Since
the coefficients $g^{(a)}_{k}$ are linear functions of the
parameter $\sigma$ the equation $c^{(a)}_{N+1}=0$ is an algebraic
equation of $(N+1)$-th order and thus it has $(N+1)$ zeros
$\sigma_{0}, \sigma_{1}, \dots, \sigma_{N}$ \cite{Ron,Fiziev6}.

In \cite{Fiziev4,Fiziev5} one can find an explicit representation of the coefficient
$c^{(a)}_{N+1}=0$ in form of a specific determinant $\Delta_{N+1}$.
This form is most convenient for practical calculations.

\subsection{Correspondence between the notations of present article and the notations,
used in computer package Maple}

The computer package Maple uses the conventions of the basic articles \cite{DDLMRR1}, \cite{DDLMRR2}.
In the Maple notation $\text{HeunC}(\alpha,\beta,\gamma,\delta,\eta,z)$ for the solution (\ref{Hca})
the parameters $\alpha,\beta,\gamma,\delta,\eta$ are related in the following way with the parameters of the
non-symmetrical canonical form of confluent Heun equation \cite{Ron}, used in the present article, too:
\begin{eqnarray}
\label{Maple} \alpha_{{}_{Maple}}=4p,\,\,\,
\beta_{{}_{Maple}}=\gamma-1,\,\,\, \gamma_{{}_{Maple}}=\delta-1,
\\ \delta_{{}_{Maple}}=4p\alpha-2p(\gamma+\delta),\,\,\,
\eta_{{}_{Maple}}=2p\gamma-{\frac{\gamma\delta-1} 2}-\sigma.
\nonumber
\end{eqnarray}
In \cite{Fiziev6,Fiziev4,Fiziev5} a modified Maple-like parametrization of the
confluent Heun equation is used:
\begin{align}\label{DHeunC}
{{d^2H}\over{dz^2}}+\left(\alpha+{{\beta+1}\over{z}}+{{\gamma+1}\over{z-1}}\right){{dH}\over{dz}}+
\left( {\mu\over z}+{\nu\over{z-1}} \right)H = 0.
\end{align}
The equation (\ref{DHeunC}) has a uniform shape.
This uniform parametrization simplifies the explicit expressions for the coefficients
in the series (\ref{series}) and (\ref{seriesinf}). For the parameters $\mu$ and $\nu$
one obtains the following relations with the parametrization, used in present article:
\begin{eqnarray}
\label{mu_nu}
\mu=\sigma,\,\,\,\mu+\nu=4p\alpha.
\nonumber
\end{eqnarray}

The first polynomial condition (\ref{alphaint}) in Maple
notations, as well as in the above uniform parametrization, reads:
$${\frac{\delta} {\alpha}}+{\frac{\beta+\gamma}{2}}+N+1=0.$$ It
yields discrete values $\delta=-\alpha\left({{1} \over
{2}}(\beta+\gamma)+N+1\right)$ of the Maple parameter $\delta$.
Hence, the name $\delta_N$-condition
\cite{Fiziev6},\cite{Fiziev4,Fiziev5}.

\end{document}